\documentstyle[twocolumn,prl,aps]{revtex}
\begin{document}
\flushbottom \draft
\title{Generating entangled atom-photon pairs from Bose-Einstein
condensates}
\author{M. G. Moore and P. Meystre}
\address{Optical Sciences Center\\
University of Arizona, Tucson, Arizona 85721\\ (April 20, 1999)
\\ \medskip}\author{\small\parbox{14.2cm}{\small \hspace*{3mm}
We propose using spontaneous Raman scattering from an optically
driven Bose-Einstein condensate as a source of atom-photon pairs
whose internal states are maximally entangled. Generating
entanglement between a particle which is easily transmitted (the
photon) and one which is easily trapped and coherently manipulated
(an ultracold atom) will prove useful for a variety of
quantum-information related applications. We analyze the type of
entangled states generated by spontaneous Raman scattering and
construct a geometry which results in maximum entanglement.
\\[3pt]PACS numbers: 03.75.Fi, 03.65.Bz, 03.67.-a }}
\maketitle \narrowtext

The nonlocal nature of quantum entanglement has remained a subject
of great interest since the earliest days of quantum mechanics
\cite{EinPodRos35}. Its transformation from the realm of gedanken
experiments to laboratory reality began in the mid 1960's, when
research involving the cascade emission of entangled photon pairs
from calcium atoms began \cite{KocCom67}. The primary motivation
for early experiments \cite{FreCla72,AspGraGer81} was to test
Bell's inequality \cite{Bel64} in order to demonstrate that
physical reality cannot be described by a local hidden variables
theory. Research involving entangled photon pairs has proceeded to
move into applications in recent years, with successes in quantum
cryptography \cite{BenBraEke92} and quantum teleportation
\cite{BenBraCre93,BouPanMat97} paving the way towards a variety of
useful quantum informational devices. In this Letter we propose a
scheme in which a Bose-Einstein condensate
\cite{AndEnsMat95,DavMewAnd95} is used to generate atom-photon
pairs whose internal degrees of freedom are similarly entangled.
This has obvious advantages for quantum informational
applications: photons are well suited for transmission over long
distances but are difficult to store at a fixed location, while
the reverse is true for atoms. Thus creating an entangled pair of
particles where one member is readily used to couple to a distant
system while the other is stored by the sender seems ideal for use
in quantum cryptography and teleportation related devices. The
development of a practical entangled atom-photon source should
also allow the first delayed choice test of Bell's inequality
involving particles other than photons.

The most widely used technique for generating entangled photon
pairs at the present time is spontaneous parametric
down-conversion in a nonlinear crystal. This system has two main
elements: a pump laser, which acts as a source of photons with
well defined energy and momentum characteristics; and a nonlinear
medium, which absorbs pump photons and spontaneously emits pairs
of photons whose internal polarization states are maximally
entangled. In recent work \cite{MooMey99a,MooZobMey99} we have
shown that the interaction between a Bose Einstein condensate and
an off-resonant light field takes the form of a cubic
nonlinearity. In this case a pump photon and a condensate atom may
be `absorbed' simultaneously and a correlated atom-photon pair
spontaneously `emitted' in a manner closely analogous to optical
parametric down-conversion. In our previous work we assumed that
only a single atomic ground state was involved, which precluded
any entanglement between the internal states of the atom-photon
pair constituents. In reality, however, the atomic species used in
BEC experiments to date have multiple ground hyperfine states,
hence a treatment of spontaneous light scattering must include the
possibility of Raman transitions between these states. Provided
that a suitable geometry can be found, this then makes it possible
to generate atom-photon pairs whose internal states (ground
hyperfine level and photon polarization) are maximally entangled.

When a sufficient feedback mechanism exists, parametric devices
such as these can operate as phase-coherent amplifiers. In the
optical case of two-photon down conversion, a resonator cavity is
typically required to achieve amplification. In the case of an
optically driven BEC, however, the scattered atoms remain in the
volume of the BEC long enough to provide feedback via Bose
enhancement, resulting in mirrorless amplification, or matter-wave
superradiance \cite{InoChiSta99,MooMey99b,LawBig98,Ino99,Koz99}.
In an experiment in which one atom-photon pair is utilized at a
time it is therefore necessary to work below the superradiance
threshold. In this regime the Bose degeneracy of the condensate
does not play a direct role in terms of providing Bose
enhancement. However, due to its high phase-space density a BEC is
still the ideal atom source for such a device.

High phase-space density is advantageous in two ways. First, due
to the subrecoil velocity spread of a condensate, an entangled
atom moving at the recoil velocity is readily resolved in momentum
space from the remaining condensate atoms. On the other hand,
because the recoil velocity is very small the spatial extent of
the atom source should be minimized so that the recoiling atom
quickly reaches an empty region of space where it can be
manipulated and/or detected, a condition which also minimizes the
occurrence of decoherence due to collisions with condensate atoms.
Taken in combination, these two requirements are best satisfied by
a minimum-uncertainty-state atom source such as a BEC
\cite{SteInoChi99}. One could also imagine creating an entangled
atom-photon pair by scattering laser light from a trapped ion or
single atom. This scheme does not seem very practical as the
probability of scattering a second photon, and therefore causing
decoherence, is comparable to the probability of scattering the
first photon. In contrast, the large number of atoms in the
initial condensate state guarantees that the probability of
scattering a single atom from the initial cloud is much larger
than the probability that the same atom scatters a second photon.
This can then lead to a steady production of entangled atom-photon
pairs, with little probability for decoherence due to multiple
scattering by the same atom.

In the remaining part of this Letter we introduce the basic
formalism to describe spontaneous Rayleigh and Raman scattering
from an optically driven BEC. We then focus on the case of a
sodium BEC and construct a specific geometry which results in the
emission of atom-photon pairs whose internal states are maximally
entangled.

In the electric-dipole and rotating-wave approximations the
atom-field interaction in first-quantized form is given by
\begin{equation}
    V_{int}=-{\bf d}^+\cdot{\bf E}^+({\bf r})-{\bf d}^-\cdot{\bf E}^-({\bf
r}),
    \label{Vint}
\end{equation}
where
\begin{equation}
    {\bf d}^+={\sum_{Fm}}^{(e)}{\sum_{F'm'}}^{(g)}|Fm \rangle
    \langle Fm|e{\bf r}|F'm'\rangle\langle F'm'|
    \label{d+}
\end{equation}
is the atomic raising operator, ${\bf E}^+({\bf r})$ is the
positive frequency part of the electric field, and ${\bf
d}^-\cdot{\bf E}^-({\bf r})= ({\bf d}^+\cdot{\bf E}^+({\bf
r}))^\dag$. The notation ${\sum}^{(e)}$ indicates a sum over
excited hyperfine sublevels, whereas ${\sum}^{(g)}$ indicates a
sum over ground hyperfine states.

Second quantization of Eq. (\ref{Vint}) with respect to both the
atomic and electromagnetic fields then gives the atom-field
interaction operator
\begin{eqnarray}
    \hat{{\cal V}}_{int}&=&-\hbar
{\sum_{Fm}}^{(e)}{\sum_{F'm'}}^{(g)}\sum_{{\bf k},\lambda}
    \int d^3r\, \hat{\mit\Psi}^\dag_e(Fm{\bf r})g_{{\bf k}\lambda}(FmF'm')
    \nonumber\\
    &\times& e^{i{\bf k}\cdot{\bf r}}\hat{b}_{{\bf k}\lambda}
    \hat{\mit\Psi}_g(F'm'{\bf r})
    +H.c.,
    \label{vint2}
\end{eqnarray}
where $\hat{\mit\Psi}_g(Fm{\bf r})$ is the annihilation operator
for a ground-state atoms with hyperfine quantum numbers $F$ and
$m$, $\hat{\mit\Psi}_e(Fm{\bf r})$ is the annihilation operator
for excited-state atoms, and $\hat{b}_{{\bf k}\lambda}$ is the
annihilation operator for photons with wave vector ${\bf k}$ and
polarization ${\bf\epsilon}_{{\bf k}\lambda}$, $\lambda=1,2$
corresponding to any two orthogonal polarization vectors for a
given ${\bf k}$. The coupling coefficient $g_{{\bf
k}\lambda}(FmF'm')$ is given by
\begin{eqnarray}
    \hbar g_{{\bf k}\lambda}(FmF'm')&=&{\cal E}_{\bf k}
    \sum_q({\bf\epsilon}^\ast_q\cdot{\bf \epsilon}_{{\bf k}\lambda})
    \langle Fm|e{\bf r}\cdot{\bf\epsilon}_q|F'm'\rangle,
    \label{gkl}
\end{eqnarray}
where ${\cal E}_{\bf k}$ is the `electric field per photon' at
wave vector ${\bf k}$ and the three unit vectors
${\bf\epsilon}_q$, $q=\{-1,0,+1\}$ are the usual polarization
vectors for $\sigma_-$, $\pi$, and $\sigma_+$ polarized photons.
They are given explicitly by ${\bf\epsilon}_{\mp 1}=\pm(\hat{\bf
x} \mp i\hat{\bf y})/\sqrt{2}$ and ${\bf\epsilon}_0 =\hat{\bf z}$.

The dipole matrix elements $\langle Fm|e{\bf r}\cdot{\bf
\epsilon}_q|F'm'\rangle$ can be decomposed in the standard way as
\begin{eqnarray}
    & &\langle Fm|e{\bf r}\cdot{\bf\epsilon}_q|F'm'\rangle=
    -(-1)^{-I-J'-F}\sqrt{(2F'+1)(2J+1)}\nonumber\\
    & &\qquad\qquad\times{\cal D}\langle Fm|F'1m'q\rangle
    \left\{\matrix{
    I&J'&F'\cr
    1&F'&J\cr}\right\},
    \label{mtrxlmnt}
\end{eqnarray}
where $I$ is the nuclear spin quantum number, $J$ and $J'$ are the
angular momentum quantum numbers of the excited and ground states
respectively,  ${\cal D}$ is the reduced dipole moment for the
$J'\leftrightarrow J$ transition, $\langle Fm|F'1m'q\rangle$ is a
Clebsch-Gordan coefficient and the quantity in brackets is a
Wigner 6-j symbol. For sodium we have $I=3/2$, $J=3/2$, and
$J'=1/2$.

The Heisenberg equations of motion for the atomic and optical
field operators are taken to be
\begin{eqnarray}
    & &\frac{d}{dt}\hat{\mit\Psi}_e(Fm{\bf
r})=-i\left[-\frac{\hbar}{2m}\nabla^2
    +\omega_e(Fm)\right]\hat{\mit\Psi}_e(Fm{\bf r})\nonumber\\
    & &\qquad+i{\sum_{F'm'}}^{(g)}\sum_{{\bf k}\lambda}g_{{\bf
k}\lambda}(FmF'm')
    e^{i{\bf k}\cdot{\bf r}}\hat{b}_{{\bf
k}\lambda}\hat{\mit\Psi}_g(F'm'{\bf r}),
    \label{dpsiedt}
\end{eqnarray}
\begin{eqnarray}
    & &\frac{d}{dt}\hat{\mit\Psi}_g(Fm{\bf
r})=-i\left[-\frac{\hbar}{2m}\nabla^2
    +\omega_g(Fm)\right]\hat{\mit\Psi}_g(Fm{\bf r})\nonumber\\
    & &\qquad+i{\sum_{F'm'}}^{(e)}\sum_{{\bf k}\lambda}g^\ast_{{\bf
k}\lambda}(F'm'Fm)
    e^{-i{\bf k}\cdot{\bf r}}\hat{b}^\dag_{{\bf
k}\lambda}\hat{\mit\Psi}_e(F'm'{\bf r}),
    \label{dpsigdt}
\end{eqnarray}
and
\begin{eqnarray}
    & &\frac{d}{dt}\hat{b}_{{\bf k}\lambda}=-i\omega_{\bf k}\hat{b}_{{\bf
k}\lambda}
    +i{\sum_{Fm}}^{(e)}{\sum_{F'm'}}^{(g)}\int d^3r\, g^\ast_{{\bf
k}\lambda}(FmF'm')\nonumber\\
    & &\qquad\times\hat{\mit\Psi}^\dag_g(F'm'{\bf r})e^{-i{\bf k}\cdot{\bf
r}}
    \hat{\mit\Psi}_e(Fm{\bf r})
    \label{dbdt}
\end{eqnarray}
where $\hbar\omega_e(Fm)$ and $\hbar\omega_g(Fm)$ are the energies
of the various hyperfine levels and $\hbar\omega_{\bf k}$ is the
energy of a photon with wave vector ${\bf k}$ . We neglect for
simplicity effects not directly related to two-photon scattering,
thus Eqs. (\ref{dpsiedt}) and (\ref{dpsigdt}) do not include
trapping potentials or collisions.

We consider the case where the system is driven by a strong
off-resonant pump laser of frequency $\omega_L$ far detuned from
any ground-excited transition, so that the excited state dynamics
can be adiabatically eliminated. With the assumption that only
photons of frequencies close to $\omega_L$ are scattered, the
adiabatic solution to Eq. (\ref{dpsiedt}) is
\begin{eqnarray}
    & &\hat{\mit\Psi}_e(Fm{\bf r})\approx-{\sum_{F'm'}}^{(g)}\sum_{{\bf
k}\lambda}
    \frac{g_{{\bf k}\lambda}(FmF'm')}{\Delta(F')}
    e^{i{\bf k}\cdot{\bf r}}\nonumber\\
    & &\qquad\times\hat{b}_{{\bf k}\lambda}\hat{\mit\Psi}_g(F'm'{\bf r}),
    \label{adelim}
\end{eqnarray}
where the detuning is given by the approximation
\begin{equation}
    \Delta(F')\approx\omega_L-\omega_a+\delta(F').
    \label{detuning}
\end{equation}
Here $\omega_a$ is the atomic resonance frequency and $\delta(F')$
is the frequency separation between ground hyperfine level $F'$
and the lowest ground hyperfine level. In deriving Eq.
({\ref{detuning}) we have neglected the difference in the kinetic
energies of the excited and ground state atomic field due to
recoil, as well as the excited state hyperfine splittings. These
latter terms are neglected under the assumption that the far
off-resonant condition $\Delta(F')\gg\Gamma$, the spontaneous
decay rate, is satisfied and that the neglected frequency shifts
are comparable to or less than $\Gamma$.

The effective equations which then govern the interaction between
the ground-state atomic field and the far off-resonant light field
are found by substituting Eq. (\ref{adelim}) in Eqs.
(\ref{dpsigdt}) and (\ref{dbdt}). For the scope of this Letter we
need only the form of the interaction operator which, together
with the self-energy terms for the atomic and optical fields,
would make up the effective Hamiltonian from which those equations
may be derived. This effective interaction operator is found to be
\begin{eqnarray}
    & &\hat{{\cal V}}_{eff}=
    {\sum_{FmF'm'}}^{(g)}\sum_{{\bf k}\lambda{\bf k}'\lambda'}
    \int d^3r\,\hat{\mit\Psi}^\dag_g(Fm{\bf r})e^{-i({\bf k}-{\bf
k}')\cdot{\bf r}}
    \nonumber\\
    & &\qquad\times \hbar G_{{\bf k}\lambda{\bf k}'\lambda'}(FmF'm')
    \hat{b}^\dag_{{\bf k}\lambda}\hat{b}_{{\bf k}'\lambda'}
    \hat{\mit\Psi}_g(F'm'{\bf r}),
    \label{Veff}
\end{eqnarray}
where we have introduced
\begin{eqnarray}
    & &G_{{\bf k}\lambda{\bf k}'\lambda'}(FmF'm')
    ={\sum_{F''m''}}^{(e)}g^\ast_{{\bf k}\lambda}(F''m''Fm)\nonumber\\
    & &\qquad\times \frac{g_{{\bf k}'\lambda'}(F''m''F'm')}{\Delta(F')}
    \label{defG}
\end{eqnarray}
as the effective coupling constant. By scattering photons with
different polarizations it is possible for an atom to jump from
one ground hyperfine sublevel into another. We now discuss how to
use these Raman transitions to generate entangled atom-photon
pairs in a controlled manner from a BEC.

The initial state of the atomic field for the case of a nearly
pure condensate is well approximated by assuming that a fixed
number of atoms $N_0$ occupy the mode $\varphi_0(Fm{\bf r})$, with
all other modes taken to be in the vacuum state. Similarly, the
initial state of the optical field can be described as consisting
of a single (pump) mode in the coherent state $\beta_L$, all other
modes being in the ground (vacuum) state. For times short enough
so that the fraction of scattered atoms is negligible we may in
addition neglect the dynamical evolution of the condensate and
pump laser, i.e., we make the undepleted pump approximation for
both the atomic and optical source fields. Since the states of the
BEC and pump laser modes are no longer treated dynamically, they
can be factored out of the Hilbert space, in which case the
interaction operator acting on the space of initially unoccupied
states is found by making the substitutions
\begin{eqnarray}
    \hat{\mit\Psi}_g(Fm{\bf r})&\to&\varphi_0(Fm{\bf r})\sqrt{N_0}
    +\hat{\psi}(Fm{\bf r}),\nonumber\\
    \hat{b}_{{\bf K}1}&\to&\beta_L,
    \label{lasers}
\end{eqnarray}
where $\hat{\psi}(Fm{\bf r})$ is the annihilation operator for
atoms in the subspace orthogonal to $\varphi_0(Fm{\bf r})$, and
${\bf K}$ is the wave vector of the pump laser.

Treating the BEC and pump laser in this manner, we now focus on
the terms in (\ref{Veff}) which describe the spontaneous
scattering of a pump photon by a condensate atom resulting in the
creation of a correlated atom photon pair. The term describing a
photon scattered into the ${\bf k}$-direction is
\begin{eqnarray}
    & &\hat{{\cal V}}_{\bf k}=\sum_\lambda{\sum_{FmF'm'}}^{(g)}
    \hbar\sqrt{N_0}\beta_L\hat{b}^\dag_{{\bf k}\lambda}\int d^3r\,
    \hat{\psi}^\dag(Fm{\bf r})\nonumber\\
    & &\qquad\times G_{{\bf k}\lambda{\bf K}_1}(FmF'm')
    e^{-i({\bf k}-{\bf K})\cdot{\bf r}}\varphi_0(F'm'{\bf r}).
    \label{Vk}
\end{eqnarray}
This can be re-expressed as
\begin{equation}
    \hat{{\cal V}}_{\bf k}=\sum_\lambda\hbar\sqrt{N_0}\beta_L
    {\cal N}_{{\bf k}\lambda}\hat{b}^\dag_{{\bf k}\lambda}\hat{c}^\dag_{{\bf
k}\lambda},
    \label{Vk2}
\end{equation}
where $\hat{c}^\dag_{{\bf k}\lambda}$ creates an atom in the state
\begin{eqnarray}
    & &\varphi_{{\bf k}\lambda}(Fm{\bf r})={\cal N}^{-1}_{{\bf k}\lambda}
    {\sum_{F'm'}}^{(g)}G_{{\bf k}\lambda{\bf K}1}
    (FmF'm')e^{-i({\bf k}-{\bf k}_L)\cdot{\bf r}}\nonumber\\
    & &\qquad\times\varphi_0(F'm'{\bf r}),
    \label{phik}
\end{eqnarray}
${\cal N}_{{\bf k}\lambda}$ being a normalization constant. The
frequency of the scattered light in a given direction is
determined from energy conservation. Taking into account the
recoil energy of the atom, we find a unique photon frequency
corresponds to each scattering direction. In case the atoms change
their ground hyperfine quantum number, one also needs to account
for the associated hyperfine splitting. This results in an
entanglement between the internal atomic state and the photon
frequency as well as its polarization.

From Eq. (\ref{Vk2}) we see that the scattering of a pump photon
in the ${\bf k}$ direction results in the creation of an
atom-photon pair in which the spinor state of the atom is
entangled with the polarization state of the photon. While in
general, the state created when $\hat{\cal V}_{\bf k}$ acts on the
vacuum is an entangled state, it is certainly possible that the
two spinors $\varphi_{{\bf k}1}(Fm{\bf r})$ and $\varphi_{{\bf
k}2}(Fm{\bf r})$ may be identical, or one may in fact be zero and
therefore non-normalizable. In either case the atom-photon state
factorizes and entanglement does not occur. We also note that the
atomic spin states corresponding to orthogonal photon
polarizations need not be themselves orthogonal. As we will see
shortly, this drawback can be eliminated by choosing an
appropriate geometry.

The specific example we consider is that of an a $F=1$, $m=0$
condensate driven by a pump laser which propagates along the
$\hat{\bf y}$ axis and is polarized along the $\hat{\bf z}$ axis.
The condensate spinor is therefore
\begin{equation}
    \varphi_0(Fm{\bf r})=\phi_0({\bf r})\delta_{F1}\delta_{m0},
    \label{MIT}
\end{equation}
and the pump polarization is given by ${\bf \epsilon}_{{\bf
K}1}={\bf \epsilon}_0$. In this case a scattered photon with
polarization ${\bf\epsilon}_{{\bf k}\lambda}$ is correlated with
an atom in state
\begin{equation}
    \varphi_{{\bf k}\lambda}(Fm{\bf r})={\cal N}^{-1}_{{\bf k}\lambda}
    G_{{\bf k}\lambda{\bf K}1}(Fm10)
     e^{-i({\bf k}-{\bf K})\cdot{\bf r}}\phi_0({\bf r}).
    \label{phi0a}
\end{equation}
The internal state $|S_{{\bf k}\lambda}\rangle$ of the scattered
atom is now completely specified by
\begin{eqnarray}
    & &\langle Fm|S_{{\bf k}\lambda}\rangle={\cal N}^{-1}_{{\bf k}\lambda}
    G_{{\bf k}\lambda{\bf K}1}(Fm10)\nonumber\\
    & &\qquad={\cal N}^{-1}_{{\bf k}\lambda}
    \frac{{\cal E}^\ast_{\bf k}{\cal E}_{{\bf K}}}{\hbar^2\Delta(1)}
    \sum_{q}{\sum_{F''m''}}^{(e)}
    ({\bf\epsilon}_q\cdot{\bf\epsilon}^\ast_{{\bf k}\lambda})\nonumber\\
    & &\qquad\times\langle F''m''|e{\bf
r}\cdot{\bf\epsilon}_q|Fm\rangle^\ast
    \langle F''m''|e{\bf r}\cdot{\bf\epsilon}_{0}|10\rangle,
    \label{sklambda}
\end{eqnarray}
which  can be evaluated explicitly by making use of the matrix
elements (\ref{mtrxlmnt}).

The pump laser consists of $\pi$-polarized light, hence the atoms
are excited from the $F=1$, $m=0$ ground state into the $F=0$,
$m=0$ and $F=2$, $m=0$ excited states states. If we then set up
our photon detector along the $\hat{\bf z}$ axis we guarantee that
only $\sigma_+$ and $\sigma_-$ photons will be detected. By taking
our basis for the scattered photon polarization to be ${\bf
\epsilon}_{{\bf z}1}={\bf \epsilon}_{+1}$ and ${\bf \epsilon_{{\bf
z}2}}={\bf \epsilon}_{-1}$ we then find that the resulting
entangled atom-photon state $|\psi_e\rangle$ is given by
\begin{eqnarray}
    & &|\psi_e\rangle=\frac{1}{2\sqrt{2}}
    \big[|\omega_1,\sigma_+\rangle|1,-1\rangle
    +\sqrt{3}|\omega_2,\sigma_+\rangle|2,-1\rangle\nonumber\\
    & &\qquad+|\omega_1,\sigma_-\rangle|1,1\rangle
    -\sqrt{3}|\omega_2,\sigma_-\rangle|2,1\rangle\big],
    \label{state1}
\end{eqnarray}
where the photon state is specified by its frequency and
polarization and the atomic hyperfine state is specified in terms
of $|F,m\rangle$ states. The frequency $\omega_1$ of photons
emitted during the $F=1 \to F=1$ atomic ground-state transition
differs from the frequency $\omega_2$ of photons emitted during
the $F=1\to F=2$ transition by the ground-state hyperfine
splitting frequency, which in the case of sodium is of the order
of GHz.

If the photodetectors are preceded by narrow spectral filters such
that only photons of the shorter wavelength are detected, then
they are entangled only with atoms in the $F=1$ state, in which
case the atom-photon state, specified by the photon polarization
and the atomic $m$ quantum number, is given by
\begin{equation}
|\psi_e\rangle=\frac{1}{\sqrt{2}}(|\sigma_+,-1\rangle+|\sigma_-,1\rangle).
    \label{state1a}
\end{equation}
Likewise, detecting only the longer-wavelength photons entangled
with $F=2$ atoms would select the antisymmetric entangled state
\begin{equation}
|\psi_e\rangle=\frac{1}{\sqrt{2}}(|\sigma_+,-1\rangle-|\sigma_-,1\rangle).
    \label{state1b}
\end{equation}

In summary, we propose a realistic scheme to generate entangled
atom-photon pairs via off-resonant light scattering from a
Bose-Einstein condensate. In view of the tremendous recent
experimental advances in the nonlinear mixing of optical and
matter waves, which has resulted in the demonstration of
matter-wave superradiance \cite{InoChiSta99} and the
phase-coherent amplification of matter waves \cite{Ino99,Koz99},
all elements are present to demonstrate this effect in the near
future. Much effort has recently been devoted to the transfer of
quantum coherence and correlations between optical and atomic
systems, so as to be able to combine the advantages of atoms as a
storage media and of light as a carrier of quantum information
\cite{MaiHagNog97,ParKim99}. We believe that the proposed scheme
is a natural and important step towards achieving this goal.

\acknowledgements

This work is supported in part by the U.S.\ Office of Naval
Research under Contract No.\ 14-91-J1205, by the National Science
Foundation under Grant No.\ PHY98-01099, by the U.S.\ Army
Research Office, and by the Joint Services Optics Program.

\end{document}